# Priming intuition disfavors instrumental harm but not impartial beneficence

Valerio Capraro[1], Jim A. C. Everett[2,3], Brian D. Earp[4]

[1]Middlesex University London, [2]Leiden University, [3]University of Kent, [4]Yale University

**Abstract**

Understanding the cognitive underpinnings of moral judgment is one of most pressing problems in psychological science. Some highly-cited studies suggest that reliance on intuition decreases utilitarian (expected welfare maximizing) judgments in sacrificial moral dilemmas in which one has to decide whether to instrumentally harm (IH) one person to save a greater number of people. However, recent work suggests that such dilemmas are limited in that they fail to capture the positive, defining core of utilitarianism: commitment to impartial beneficence (IB). Accordingly, a new two-dimensional model of utilitarian judgment has been proposed that distinguishes IH and IB components. The role of intuition on this new model has not been studied. Does relying on intuition disfavor utilitarian choices only along the dimension of instrumental harm or does it also do so along the dimension of impartial beneficence? To answer this question, we conducted three studies (total N = 970, two preregistered) using conceptual priming of intuition versus deliberation on moral judgments. Our evidence converges on an interaction effect, with intuition decreasing utilitarian judgments in IH—as suggested by previous work—but failing to do so in IB. These findings bolster the recently proposed two-dimensional model of utilitarian moral judgment, and point to new avenues for future research.

**Introduction**

Understanding how ordinary people make decisions within the moral domain is of profound importance, both theoretically and practically. Theoretically, it is important because of the fundamental role that moral considerations play in numerous aspects of our cognitive life: if we want to fully understand how the mind works, we must understand the moral mind. Practically, it is important because we are all affected by the moral decisions of those around us, often in serious ways: effective public policy, for instance, depends on a keen appreciation of how the moral psychology of ordinary people actually works. It is no surprise, then, that interest in moral psychology has blossomed in the last few decades.

Over those years, work in the area has been dominated by the use of what might be called 'sacrificial' moral dilemmas. These refer to—usually hypothetical—situations in which a person must decide whether to endorse an action that is expected to maximize welfare (e.g., save the most number of lives) while foreseeably causing the death of at least one innocent person, often instrumentally. For example, is it morally permissible to torture an innocent person to death if this would be necessary to prevent a major terrorist attack that would kill hundreds of people, assuming that no one would find out about the torture? Most people recognize that there is a tension between two competing moral positions in such cases, but different people may resolve this tension differently. In the terrorism example, those who endorse torturing the innocent person are typically said to be making a 'consequentialist' judgment, because in this specific context such an action appears to be consistent with what is required by moral theories holding that the rightness or wrongness of an action depends only on its consequences. A particularly famous consequentialist theory is act utilitarianism, which holds, more specifically, that "actions are right in proportion as they tend to promote happiness, wrong as they tend to produce the reverse of happiness" (Mill, 1863; see also Bentham 1789/1983). In line with the focus of most other work in moral psychology, we will concern ourselves only with act utilitarianism in this paper, setting aside other consequentialist theories (e.g., rule-based theories, or theories holding that the moral status of an action depends on consequences other than happiness or well-being).

There are many ways to reject utilitarianism—for example, one might be inclined toward Aristotelian virtue ethics, or a feminist ethics of care—but the main non-utilitarian moral theory discussed in the contemporary moral psychology literature is 'deontology.' Broadly speaking, a deontological moral theory holds that the rightness or wrongness of an action depends on whether it fulfils certain moral norms, rules, or duties, regardless of the consequences (e.g. Kant, 1797/2002). Because deontology is usually treated as the main, or perhaps the only, alternative to utilitarianism, people who decline to endorse the ostensibly utilitarian option in a sacrificial moral dilemma are often said to have made a 'deontological' judgment (Greene, 2015). So, for example, if someone declines to endorse the torturing of an innocent person in the terrorism example, despite the fact that this has been stipulated to lead to the deaths of hundreds of other innocent people, it is typically assumed that this person's motivation or reasoning must be based in deontological considerations—for example, a Kantian prohibition on using other people as a mere means to an ends, or perhaps more simply, an intuitive application of the commonsense moral rule that killing innocent people is wrong (even if it may have good consequences).

According to now classic work in moral psychology (beginning with Greene, Sommerville, Nystrom, Darley & Cohen, 2001), tendencies to favour 'utilitarian' or 'deontological' resolutions to sacrificial moral dilemmas reflect two distinct and dissociable underlying cognitive processes in the psychology of ordinary people, characterized by Greene (2008) as *psychological natural kinds*. According to this view, utilitarian tendencies and deontological tendencies map onto even more basic cognitive systems that operate quite differently. One system (System 1) is said to be fast, intuitive, and primarily affective, whereas the other system (System 2) is said to be slow, deliberative, and rational (Chaiken & Trope, 1999; Epstein, 1994; Evans & Stanovich, 2013; Kahneman, 2011; Sloman, 1996; but see Melnikoff & Bargh, 2018).

Such dual-process theories have fruitfully modelled people's behaviour in a number of contexts, including problem solving (Fetterman & Robinson, 2013), consumer behavior (Shiv & Fedorikhin, 1999), person perception (Tamir, Thornton, Contreras, & Mitchell, 2016), cooperative behaviour (Rand, Greene & Nowak, 2012), altruistic behaviour (Rand et al, 2016), honest behaviour (Capraro, 2017), and, indeed, moral judgments in sacrificial dilemmas (Li et al, 2018). According to Greene's influential dual process model, 'deontological' judgments (refusing to sacrifice the one innocent person) are based in immediate intuitions or emotional gut-reactions, whereas 'utilitarian' judgments (sacrificing the innocent person to save a greater number) are uniquely attributable to effortful reasoning.

There is now a large body of evidence supporting this perspective: that is, the view that deliberation favors 'utilitarian' judgments whereas intuition favors 'deontological' judgments. (Ciaramelli, Muccioli, Ladavas & di Pellegrino, 2007; Conway & Gawronski, 2013; Cummins & Cummins, 2012; Koenigs et al, 2007; Kvaran, Nichols & Sanfey, 2013; Greene et al, 2001; Greene et al, 2008; Li, Xia, Wu & Chen, 2018; Mendez, Anderson, & Shapria, 2005; Patil et al, 2018; Spears, Fernández-Linsenbarth, Okan, Ruz & González, 2018; Suter & Hertwig, 2011; Timmons & Byrne, 2018 Trémolière & Bonnefon, 2014; Trémolière, De Neys & Bonnefon, 2012). For example, participants typically take longer to make pro-sacrifice (utilitarian) decisions, which is thought to reflect greater cognitive effort (Greene et al, 2001), whereas forcing participants to respond quickly under time pressure (Suter & Hertwig, 2011; Trémolière & Bonnefon, 2014), or increasing cognitive load (Trémolière & Bonnefon, 2014) tends to reduce the incidence of such decisions. Based upon these and similar findings, Greene and colleagues have proposed that utilitarian psychological tendencies—and even normative utilitarian philosophical theories—are rooted in higher-level, deliberative mental processes corresponding to superior moral judgment, whereas deontological psychological tendencies and associated moral theories are rooted in lower-level, emotionally-driven mental processes corresponding to unreflective gut responses (for extensive criticism of this view, see Berker, 2009).

In recent years, Kahane and colleagues (Kahane, 2015; Kahane et al, 2015; Kahane et al, 2018; Kahane & Shackel, 2010) have challenged the dual process model on the grounds that previous research has focused almost entirely on sacrificial dilemmas. According to these critics, sacrificial dilemmas are limited in that they bear on just one dimension of utilitarianism, namely, the permissibility of causing instrumental harm (IH), whereas they fail entirely to capture a second, more fundamental dimension of utilitarianism, namely, a commitment to impartial beneficence (IB). This refers to the moral requirement that one must strive to promote the greater good of all human beings (or even all sentient life) in a radically impartial way, that is, without regard to the physical, emotional, or relational distance between the actor and the beneficiary (Singer, 1979). Definitionally, such a drive to maximize the good is what utilitarianism is all about. Now sometimes, it may be the case that stringently pursuing this more basic, beneficent aim will require that one causes instrumental harm—but that is not typical case in real life. Under ordinary circumstances, impartially promoting the good of all tends to involve the very *opposite* of causing harm, namely helping others and performing good deeds. Indeed, even the most committed utilitarian would prefer to avoid causing harm, instrumental or otherwise, insofar as this could be reconciled with maximizing welfare. Therefore, the widespread focus on such harm in moral psychology studies is arguably both peculiar and misleading.

Motivated by this insight, Kahane, Everett et al. (2018) introduced a two-dimensional model of utilitarian psychology with both IH and IB components. To measure people's position in this two-dimensional space, Kahane and colleagues created, refined, and validated a new scale: the "Oxford Utilitarianism Scale" (OUS). This scale consists of nine short statements or scenarios, five in the dimension of impartial beneficence (IB) and four in the dimension of instrumental harm (IH). Kahane and colleagues found that these two dimensions are psychometrically independent, suggesting that IB and IH are dissociable not just conceptually but also psychologically. For example, empathic concern (Davies, 1980), identification with all of humanity (McFarland, Webb & Brown, 2012), and concern for future generations were found to be positively associated with IB but negatively associated with IH. Moreover, IH was correlated with subclinical psychopathy (Levenson, Kiehl, &Fitzpatrick, 1995), whereas IB was correlated with religiosity (Huber & Huber, 2012). Thus, these two dimensions have different individual correlates. Moreover, they have different second-order effects on social judgment. A number of studies have now reported that in the domain of IH, non-utilitarian agents are consistently rated as more moral and trustworthy than utilitarian agents (Bostyn & Roets, 2017; Brown & Sacco, 2017; Capraro et al, 2018; Kreps & Monin, 2014; Everett, Pizarro & Crockett, 2016; Lee, Sul & Kim, in press; Rom, Weiss & Conway, 2017; Rom & Conway, 2018; Sacco et al, 2016; Uhlmann, Zhu & Tannenbaum, 2013). This is not the case, however, within the domain of IB: Everett et al. (2018) have found that non-utilitarian agents tend to be preferred to utilitarian ones only for close interpersonal relationships (e.g., friend, spouse), but not for more distant roles (e.g., political leader).

The emerging picture thus seems to be that utilitarian decisional tendencies among everyday people do not constitute a single psychological dimension driven by deliberation, in contrast to deontological decisional tendencies driven by intuition. Rather, such utilitarian thinking appears to be itself divided into two, even more basic psychological dimensions, namely, a relative willingness to endorse—or a lack of aversion to—causing instrumental harm (IH), and a relative commitment to impartial beneficence (IB). If intuitive, System 1 mental processing is thought to disfavor utilitarian judgment—as prior research strongly suggests—we must therefore ask: Along which dimension? Does it do so just along IH, consistent with the focus of the sacrificial dilemmas current predominating in this area of research? Or does it do so also along IB, which has only recently been identified as a distinct psychological component of utilitarian thinking?

In this paper we investigate this question, reporting the results of three empirical studies (total N = 970, two preregistered). In these studies, we manipulated participants' cognitive process through conceptual priming of intuition (Shenhav, Rand & Greene, 2011; Rand, Greene, Nowak, 2012; Levine et al, 2018) and assessed their endorsement of IH and IB on the OUS. This is novel in two key ways. First and foremost, it is the first study of its kind to address the cognitive underpinnings of impartial beneficence, which we have suggested is a fundamental aspect of utilitarian judgment. Second, even within the domain of instrumental harm—which has been the focus of previous work—our method allows for a better test of the claims of the dual-process model. This is because we rely on short items from the OUS that have been extensively validated, instead of the more complicated, messy, and less well-validated sacrificial moral dilemmas (Bauman, McGraw, Bartels, & Warren, 2014). If the effect on IH can be conceptually replicated using these short, validated items, this would allow us to have greater confidence in the claims of the dual-process model regarding the IH component of utilitarian psychology.

A final contribution of this work is that it assesses, for the first time, the susceptibility of the OUS to priming. Since the OUS was introduced as a trait-level measure of individual differences in proto-utilitarian psychological tendencies, it is important to determine whether or to what extent participants' responses to the short items can be influenced by explicit instructions to rely on intuition versus deliberation.

**Study 1**

Our first experiment was a non-preregistered exploratory study looking at the effect of priming intuition versus deliberation on participants' scores on the OUS. For this and the other studies, we report all measures, manipulations and exclusions; in all the studies, data collection was not continued after the analysis.

*Method*

Participants were recruited on Amazon MTurk and paid $0.50 for participating. After eliminating duplicate IP addresses and MTurk IDs (10 observations) and/or individuals who left the OUS incomplete (1 observation), we had a final sample of 263 participants (47% females; mean age = 36.9, SD = 12.2). No a priori power analysis was conducted for this study. Sample size was determined by the amount of available funding left over from a previous project. A sensitivity power analysis based on a significance level $\alpha = 0.05$ and power $\beta = 0.8$ shows that the sample size we achieved was sufficient to detect a small effect size of $f = 0.09$. In the study, participants were randomly assigned to one of two between-subjects conditions (intuition vs. deliberation). Following Levine et al. (2018), Study 3, participants were encouraged to use their intuitive (or deliberative) system through a conceptual prime making salient how emotion (or reason) leads to "good decision making" and "satisfying decisions."

> *Conceptual priming of intuition in Study 1*
>
> *Sometimes people make decisions by using feeling and relying on their emotion. Other times, people make decisions by using logic and relying on their reason.*
>
> *Many people believe that emotion leads to good decision-making. When we use feelings, rather than logic, we make emotionally satisfying decisions. Please answer the following nine questions by relying on emotion, rather than reason.*
>
> *Conceptual priming of deliberation in Study 1*
>
> *Sometimes people make decisions by using logic and relying on their reason. Other times, people make decisions by using feeling and relying on their emotion.*
>
> *Many people believe that reason leads to good decision-making. When we use*

*logic, rather than feelings, we make rationally satisfying decisions. Please answer the following nine questions by relying on reason, rather than emotion.*

Our dependent measure was the Oxford Utilitarianism Scale (OUS), consisting of 9 items in two sub-scales to which participants indicated their agreement on a 7-point scale (*1 = strongly disagree, 7 = strongly agree*). The first subscale - Impartial Beneficence (OUS-IB) - consists of 5 items reflecting endorsement of the impartial maximization of the greater good, even at the cost of personal self-sacrifice (e.g., "If the only way to save another person's life during an emergency is to sacrifice one's own leg, then one is morally required to make this sacrifice"). The second subscale - Instrumental Harm (OUS-IH) - consists of 4 items reflecting a relative willingness to cause harm in order to bring about the greater good (e.g., "It is morally right to harm an innocent person if harming them is a necessary means to helping several other innocent people"). Mean scores on both dimensions were computed for all participants, and showed good reliability (Cronbach's $\alpha$ = 0.781 in the intuition condition, and 0.783 in the reason condition for OUS-IB; $\alpha$ = 0.726 in the intuition condition, and 0.826 in the reason condition for OUS-IH). When completing these questions, participants were reminded to "rely on emotion [reason]." Exact experimental instructions are reported in the Appendix.

To analyze results, we used linear regression, entering conceptual prime condition as a between-subjects factor (*0 = intuition, 1 = deliberation*) and dummy-coding scores on each OUS dimension as a within-subjects variable (*0 = IH; 1 = IB*)

*Results*

A linear regression predicting OUS scores as a function of conceptual prime condition, OUS dimension, and their interaction, revealed a significant overall effect ($R^2$ = 0.04, $F(3,522)$=6.84, $p<.001$). Moving to the main effects, we found a significant effect of OUS dimension ($\beta$=0.601, $t$=3.60, $p <.001$), no significant main effect of conceptual prime ($\beta$=0.123, $t$=0.68, $p <.495$), and, crucially, a significant interaction of prime and scores on each dimension ($\beta$= 0.851, $t$ = 3.38, $p < .001$). This pattern of results suggests that conceptual priming had a different impact on the two dimensions of utilitarian psychology. Breaking the interaction down by looking at each dimension separately, we replicated previous work by showing that endorsement of instrumental harm was significantly higher, $R^2$ = 0.06, $F(1,261)$=16.65, $p<.001$, when deliberation was primed (M = 4.27, SD = 1.76) than when intuition was primed (M = 3.54, SD = 1.18). Whereas by contrast, we found no significant difference, $R^2$ < 0.01, $F(1,261)$=0.47, p=4.94, in endorsement of impartial beneficence depending on whether deliberation (M = 4.02, SD = 1.44) or intuition was primed (M = 4.14, SD = 1.41). Since we had sufficient power to detect even a small effect size, we take this null finding to be meaningful (see Figure 1).

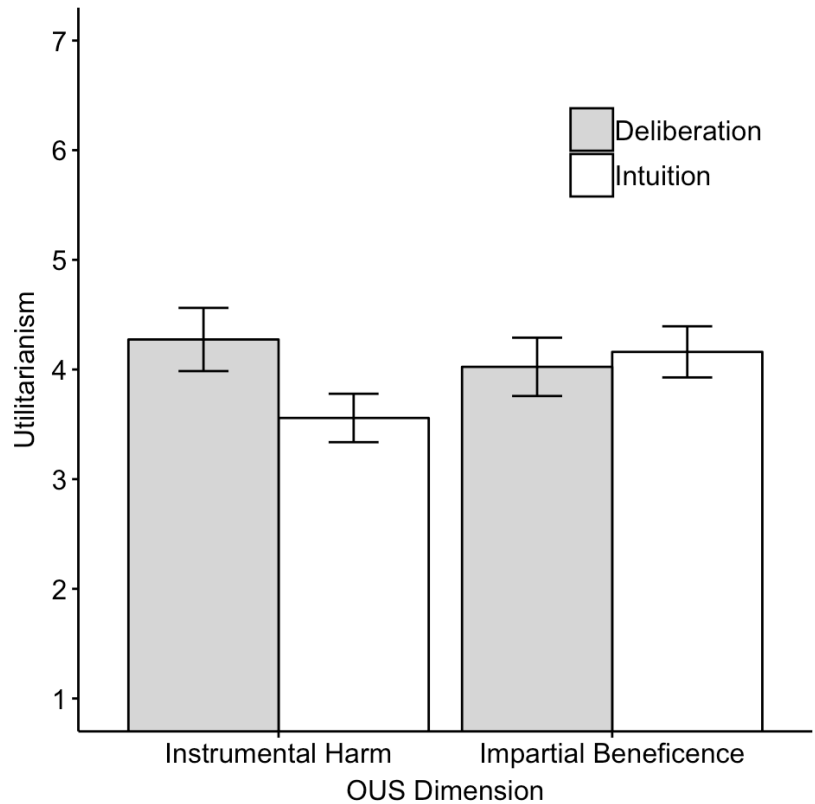

*Figure 1. Priming intuition decreases utilitarian judgments in the domain of instrumental harm, but not impartial beneficence. Error bars represent 95% confidence intervals. (Study 1).*

**Discussion**

Our first study provides initial evidence that promoting intuition decreases utilitarian judgments in the domain of instrumental harm but not in the domain of impartial beneficence. To see whether this exploratory result could be replicated and thus enhance confidence in these results, we conducted a second, pre-registered, study.

**Study 2**

The goal of the second study was to replicate the finding from Study 1 with a pre-registered design, adjusting the priming materials to more closely align with our research questions. In our context, we are not really interested in "emotionally satisfying" versus "rationally satisfying" decisions. Rather, we are interested in what people perceive to be the right thing to do. Therefore, we decided to replace, in the original primes from Levine et al. (2018) that we used in Study 1, the words "emotionally [rationally] satisfying decisions" with "better decisions," and to add one sentence that more explicitly refers to the rightness of using the positively primed cognitive system. Additionally, we decided to use the word "intuition" instead of the word "emotion," given that previous work has focused not just on the importance of emotions to non-utilitarian judgments, but also the role of intuitions more generally.

*Method*

Participants were recruited on Amazon MTurk and paid $0.50 for participating. As noted in our pre-registration, sample size was determined through an a priori power analysis showing that we would need at least 200 participants to detect a small effect size of f = 0.10, taking a significance level α=.05 and a power β = 0.80. To account for any exclusions or technical problems, we recruited 250 participants. After eliminating duplicate IP addresses and MTurk IDs (4 observations) and/or individuals who left the OUS incomplete (0 observations), we had a final sample of 246 participants (43% females; mean age = 33.8, SD = 9.9). None of these participants had participated in the previous study. A sensitivity power analysis with significance α = 0.05 and power β = 0.8 showed that our sample size was sufficient to detect a small effect of f = 0.089. The dependent variables were the same as in Study 1. The design, analysis, exclusion criteria, and sample size were pre-registered at: https://aspredicted.org/gr68s.pdf.

The design was identical to Study 1, except for the conceptual primes, which in this case were as follows:

> <u>*Conceptual priming of intuition in Study 2*</u>
>
> *Sometimes people make decisions by using feeling and relying on their intuitions. Other times, they make decisions by using logic and relying on their reason.*
>
> *Many people believe that intuition leads to good decision-making: whether something 'feels right' is often a good indication of whether it is right. When we rely on our automatic 'gut feelings', instead of just logic, we often make better decisions.*
>
> *Please answer the following nine questions by relying on your intuitions, rather than reason. When you read each question, focus on your first, emotional response and your 'gut-feeling'. Try not to think too much about each question, and instead just focus on what your intuition tells you.*
>
> <u>*Conceptual priming of deliberation in Study 2*</u>
>
> *Sometimes people make decisions by using logic and relying on their reason. Other times, they make decisions by using feeling and relying on their intuitions.*
>
> *Many people believe that reason leads to good decision-making: whether something is rational and makes logical sense is often a good indication of whether it is right. When we think carefully through a problem, rather than just going on automatic 'gut-feelings', we often make better decisions.*
>
> *Please answer the following nine questions by relying on reason, rather than intuition. When you read each question, focus on thinking and reasoning about*

> *the question. Try not to focus on what your emotional gut-reactions tell you, and instead think carefully about each question.*

*Results*

A linear regression predicting OUS scores as a function of conceptual prime condition, OUS dimension, and their interaction, revealed a marginally significant overall effect ($R^2 = 0.02$, $F(3,488)=2.47$, $p=0.06$). Moving to the main effects, we found no significant effect of OUS dimension ($β=0.238$, $t=1.34$, $p=0.181$), nor a significant main effect of conceptual prime ($β=-0.122$, $t=0.68$, $p <.495$), but a nearly significant interaction of prime and scores on each dimension, ($β = 0.337$, $t = 1.78$, $p = 0.076$). Although the observed p value did not satisfy our pre-registered alpha criterion, it was very close to it, suggesting that conceptual priming may have had a meaningfully different impact on the two dimensions of utilitarian psychology. Breaking the interaction down by looking at each dimension separately, we replicated Study 1 by showing that endorsement of instrumental harm was significantly higher, $R^2 = 0.01$, $F(1,244)=6.39$, $p=0.012$, when deliberation was primed ($M = 4.45$, $SD = 1.37$) than when intuition was primed ($M = 4.00$, $SD = 1.47$). Also consistent with Study 1, there was no significant difference, $R^2 < 0.01$, $F(1,244)=0.49$, $p=0.487$, in endorsement of impartial beneficence depending on whether deliberation ($M = 4.36$, $SD = 1.37$) or intuition was primed ($M = 4.23$, $SD = 1.35$) (see Figure 2).

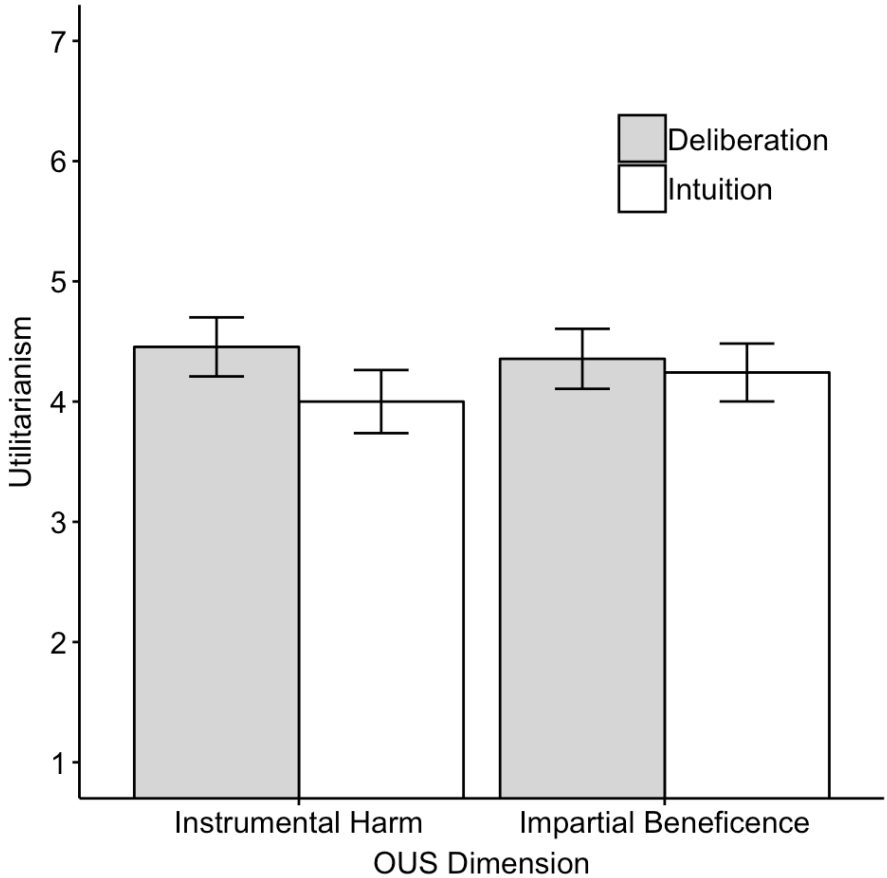

*Figure 2. Priming intuition decreases utilitarian judgments in the domain of instrumental harm, but not impartial beneficence. Error bars represent 95% confidence intervals. (Study 2).*

**Discussion**

The results of Study 2 were essentially consistent with those of Study 1, albeit with slightly different priming materials designed to more closely bear on the research question. However, the predicted interaction effect in this study was - in contrast to Study 1 - only marginally significant by conventional standards ($p = .076$). Therefore, we decided to conduct a third study, with the goal of addressing some of the potential shortcomings of Study 2.

**Study 3**

We could see three possible limitations to Study 2: perhaps, despite our power analyses, we just did not have enough statistical power for this specific effect; perhaps we used a too long a conceptual priming passage, which may have caused some participants to lose interest; and perhaps the word "intuition" was less evocative than the word "emotion." Therefore, in Study 3, we collected a larger sample size and used a conceptual prime that combined the relevant features from Study 1 and Study 2. Specifically, as in Study 1, the priming passages in Study 3 were very short and used the word "emotion" instead of the word "intuition." But as in Study 2, the priming material in Study 3 does not use the words "emotionally [rationally] satisfying decisions," but rather, "better decisions."

*Method*

Participants were recruited on Amazon MTurk and paid $0.50 for participating. After eliminating duplicate IP addresses and MTurk IDs (6 observations) and/or individuals who left the OUS incomplete (1 observation), we had a final sample of 461 participants (40% females; mean age = 35.6, SD = 10.1). As noted in our pre-registration, our sample size was determined through recruiting as many participants as we could within our budgetary constraint. Nonetheless, a sensitivity power analysis with significance $\alpha = 0.05$ and power $\beta = 0.8$ shows that our sample size was sufficient to detect a small effect of $f = 0.065$.

The design, analysis, exclusion criteria, and sample size were pre-registered at: https://aspredicted.org/9hv38.pdf. The variables were the same as in previous studies. Please note that the elimination from the analysis of the one participant who left the OUS incomplete was not pre-registered. Regardless, results remain the same if we include this participant in the analysis. None of our participants had participated in the previous studies.

The design was identical to Study 1, except for the slight differences to the conceptual primes, which were in this case as follows:

> *Conceptual priming of intuition in Study 3*
>
> *Sometimes people make decisions by using feeling and relying on their emotion. Other times, people make decisions by using logic and relying on their reason.*
>
> *Many people believe that emotion leads to good decision-making. When we use feelings, rather than logic, we make better decisions. Please answer the following nine questions by relying on emotion, rather than reason.*
>
> *Conceptual priming of deliberation in Study 3*
>
> *Sometimes people make decisions by using logic and relying on their reason. Other times, people make decisions by using feeling and relying on their emotion.*
>
> *Many people believe that reason leads to good decision-making. When we use logic, rather than feelings, we make better decisions. Please answer the following nine questions by relying on reason, rather than emotion.*

*Results*

A linear regression predicting OUS scores as a function of conceptual prime condition, OUS dimension, and their interaction revealed a marginally significant overall effect ($R^2 = 0.02$, $F(3,918)=5.65$, $p<0.001$). Moving to the main effects, we found a significant effect of OUS dimension ($\beta=0.428$, $t=3.39$, $p=0.001$), no significant main effect of conceptual prime ($\beta=0.008$, $t=0.68$, $p <.495$), and a significant interaction of prime and scores on each dimension, $\beta = 0.475$,

t = 3.53, p < .001). Consistent with Studies 1 and 2, this pattern of results suggests that conceptual priming had a different impact on the two dimensions of utilitarian psychology. Breaking the interaction down by looking at each dimension separately, we replicated Study 1 by showing that endorsement of instrumental harm was significantly higher, $R^2 = 0.02$, $F(1,459)=10.81$, $p=0.001$, when deliberation was primed (M = 4.28, SD = 1.56 than when intuition was primed (M = 3.82, SD = 1.47). In contrast, there was no statistical difference, $R^2 < 0.01$, $F(1,459)<0.01$, $p=0.949$, in endorsement of impartial beneficence depending on whether deliberation (M = 4.28, SD = 1.50) or intuition was primed (M = 4.29, SD = 1.38) (see Figure 3).

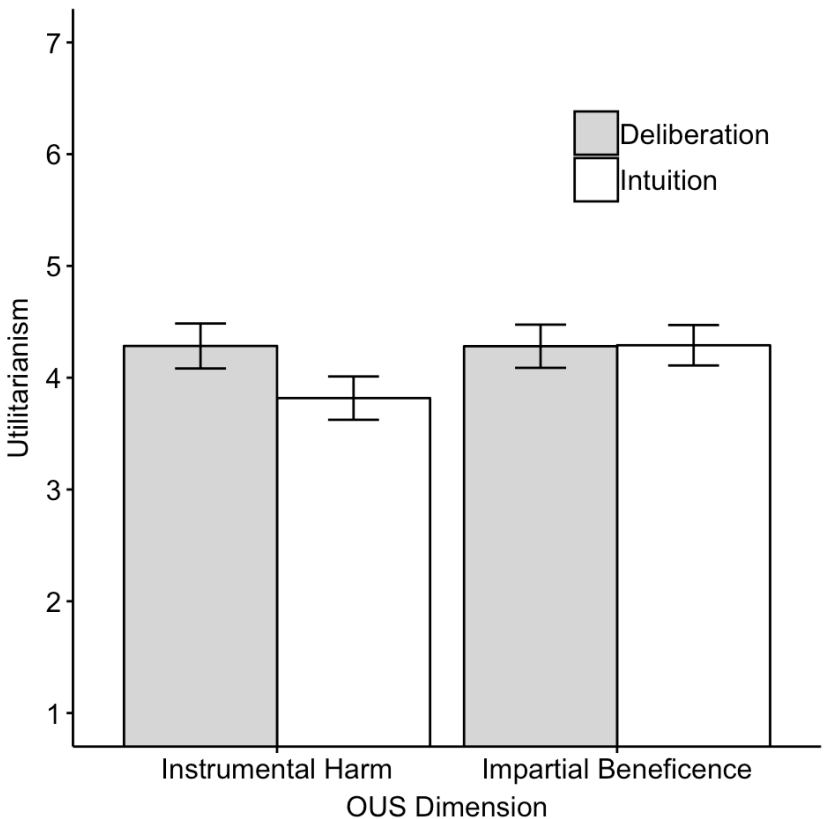

*Figure 3. Priming intuition decreases utilitarian judgments in the domain of instrumental harm, but not impartial beneficence. Error bars represent 95% confidence intervals. (Study 3).*

**Discussion**

In this final replication, we found stronger results consistent with Study 1. Across all three studies, we observed the same pattern of results, suggesting that the basic effect is real and reliable.

**General Discussion**

In the last two decades, much work in moral psychology has applied dual process models to the study of ostensibly utilitarian judgments in sacrificial moral dilemmas, concluding that non-utilitarian or 'deontological' judgments (refusing to sacrifice the one) are based in immediate intuitions and emotional gut-reactions, whereas utilitarian judgments (sacrificing the one to save a greater number) are more attributable to effortful reasoning (Ciaramelli et al, 2007; Koenigs et al, 2007; Greene et al, 2001; Greene et al, 2008; Mendez et al, 2005; Suter & Hertwig, 2011 Trémolière & Bonnefon, 2014; Trémolière et al, 2012). In recent years, however, this work has been challenged by Kahane, Everett, and colleagues (2018) who argued that proto-utilitarian decision making breaks down into a two-dimensional psychological space, and that solely studying sacrificial dilemmas will not tell us much about utilitarian psychology generally (Kahane, 2015; Kahane & Shackel, 2010).

In particular, the two-dimensional model of utilitarian psychology posits that there are at least two dimensions to consider – impartial beneficence (IB) and instrumental harm (IH) – and that these two dimensions are not just dissociable theoretically, but also empirically. These two dimensions have distinct psychological correlates (Kahane et al 2015, 2018) and even have divergent second-order effects on social perception (Everett et al, 2018). Based on a large body of work looking at the effects of intuition and deliberation on sacrificial decisions relating to IH, researchers have sought to draw conclusions about the nature of utilitarian psychology. But what about the second, more fundamental, dimension of utilitarianism – IB? In this study we investigated the role of deliberative versus intuitive cognitive processes in encouraging utilitarian decisions in both the IH and IB domains of utilitarianism.

We conducted three studies (total N = 970, two preregistered) in which we used conceptual priming to encourage participants to rely on intuition or deliberation when answering nine short questions tapping endorsement of IB or IH on the OUS. In the domain of IH, we conceptually replicated previous findings by showing that priming intuition reduces utilitarian decisions that involve causing harm for the greater good. In doing so, we demonstrate a causal link between reliance on intuition and decreased utilitarian judgments in sacrificial dilemmas using a different cognitive manipulation and different dependent measures from those used in previous studies. Specifically, we find that the effect of promoting intuition on the IH dimension of the OUS is similar to the effect of promoting intuition in classical sacrificial moral dilemmas, suggesting that the cognitive processes underlying responses to the OUS are similar to those underlying the more concrete dilemmas that have been employed in prior work. That being said, we cannot compare our results with previous findings along the IB dimension, since ours is the first set of studies to explore this dimension using current methods.

In this dimension, we report a different pattern of results to the one observed for IH. Across the three studies, while priming intuition did decrease utilitarian judgments involving instrumental harm, it failed to do so for utilitarian decisions relating to impartial beneficence. The mean judgment across all three studies in the IB dimension when deliberation was primed is very similar to the average judgment when intuition was primed (4.23 vs 4.22), suggesting that the priming effect along the IB dimension is either zero or too small to be of interest (Lakens, 2017). This different pattern of results between IH and IB suggests that these dimensions are not only psychologically distinct in terms of trait measurement, as suggested by Kahane et al. (2018), but distinct in terms of cognitive processing: conceptual priming of intuition selectively interferes with the IH dimension, but not with the IB dimension. Rather than intuition leading to non-utilitarian decisions in general, intuition seems to favor a refusal to inflict harm for the greater good specifically.

Our work has several limitations. The first one is that it focuses on judgments in hypothetical dilemmas, rather than behaviour in real-life dilemmas. Virtually all studies in moral philosophy and moral psychology have focused on hypothetical dilemmas, presumably because of the ethical difficulties that would be associated with causing real-world harm in a laboratory setting. However, a recent study by Bostyn, Sevenhant and Roets (2018) made an important step in addressing this issue by having humans making moral decisions ostensibly affecting rats rather than other humans. In their experiments, Bostyn et al. (2018) asked human participants to decide whether to administer an electroshock to one rat in order to save five rats from receiving the shock. Although in reality the shock was bogus, such that no rats were actually harmed, the authors found that participants' judgments in overtly hypothetical dilemmas were not predictive of their behaviour in these (apparently) real dilemmas. Given this discrepancy between responses in hypothetical dilemmas and actual behaviour, an important direction for future research is to explore the extent to which intuition and deliberation underlie decisions in real-life moral dilemmas.

Another limitation of our studies has to do with the use of conceptual primes. A possible downside of these primes is that they are transparent with respect to their purpose, as they explicitly mention the words emotion, intuition, and reason. This might create demand effects such that participants respond according to what they believe must be the function of intuition or reason (Rand, 2016). However, previous work suggests that conceptual primes have a similar effect as other cognitive manipulations, such as time pressure, cognitive load, and ego depletion, in several domains, including belief in God (Shenhav et al, 2012), cooperation (Rand et al, 2012), lying (Cappelen, Sørensen & Tungodden, 2013), and, indeed, moral judgments in sacrificial dilemmas (Li et al, 2018). Moreover, our specific conceptual primes led to similar results as those of other cognitive manipulations in the context of sacrificial moral dilemmas (IH dimension), suggesting that our manipulation worked in a similar fashion to other, less explicit methods of inducing a reliance on intuition. Nevertheless, future work should certainly examine the robustness of our findings across a range of cognitive manipulations.[1]

A third limitation of our studies, which applies equally the previous studies using sacrificial dilemmas, is that they cannot reveal the underlying motivations behind the responses of participants. Why does intuition promote non-utilitarian judgments in the IH dimension, but not in the IB dimension? At this stage, we can only speculate. One view is that deontological rules came about to function as simple heuristics that work well to promote human flourishing in most ordinary circumstances (Baron, 1994; Greene, 2008), but which must be overridden in special circumstances to achieve the same goal. Given that, psychologically, "bad is stronger than good" (Baumeister, Bratslavsky, Finkenauer, & Vohs, 2001), heuristics forbidding causing harm are likely to be more automatic and intuively accessible than heuristics about non-obligatory help. Another potential explanation for why intuition does not affect moral judgments in the IB dimension might be that the kind of partiality (e.g., favoring friends and family) that characterizes the decisions of those low in IB would have been highly adaptive for our ancestors, and thus robust against competing deliberations (Bloom, 2011). Exploring these and other possible motivations that could underlie our effect should be a focus of subsequent studies.

---

[1] A related question concerns whether participants can meta-cognitively direct themselves to think intuitively or deliberately. This could be tested, for example, by asking participants to imagine some other participant answering the OUS after viewing the primes (or under some other cognitive manipulation), and assessing whether they make correct predictions about the responses of these participants. This is certainly a promising direction for future research that would go beyond the specific domain of moral judgments in the OUS: a similar question can be asked with respect to other cognitive processes that have been shown to affect people's decisions, such as cooperation (Rand, 2016), altruism (Rand et al, 2016), and negative reciprocity (Hallsson, Hartwig & Hulme, 2018).

To conclude, we report that conceptual priming of intuition decreases endorsement of instrumental harm but not impartial beneficence. This finding adds to work suggesting that utilitarian thinking cannot be understood solely in terms of psychological states or processes associated with a greater willingness to cause instrumental harm. Instead, it will be important to continue study the IH and IB dimensions of utilitarian psychology as separate constructs going forward.

# Appendix

We report the experimental instructions of Study 1. Those of Study 2 and Study 3 are identical, apart from the primes, as described in the main text.

*Experimental instructions of Study 1, Priming Emotion condition*

Sometimes people make decisions by using feeling and relying on their emotion. Other times, people make decisions by using logic and relying on their reason.

Many people believe that emotion leads to good decision-making. When we use feelings, rather than logic, we make emotionally satisfying decisions. Please answer the following nine questions by relying on emotion, rather than reason.

(page break)

"If the only way to save another person's life during an emergency is to sacrifice one's own leg, then one is morally required to make this sacrifice."

Rely on emotion.

*(answers collected using a 7-item Likert scale from "Strongly Disagree" to "Strongly Agree")*

(page break)

"It is morally right to harm an innocent person if harming them is a necessary means to helping several other innocent people."

Rely on emotion.

*(answers collected using a 7-item Likert scale from "Strongly Disagree" to "Strongly Agree")*

(page break)

"From a moral point of view, we should feel obliged to give one of our kidneys to a person with kidney failure since we don't need two kidneys to survive, but really only one to be healthy."

Rely on emotion.

*(answers collected using a 7-item Likert scale from "Strongly Disagree" to "Strongly Agree")*

(page break)

"If the only way to ensure the overall well-being and happiness of the people is through the use of political oppression for a short, limited period, then political oppression should be used."

Rely on emotion.

*(answers collected using a 7-item Likert scale from "Strongly Disagree" to "Strongly Agree")*

(page break)

"From a moral perspective, people should care about the well-being of all human beings on the planet equally; they should not favor the well-being of people who are especially close to them either physically or emotionally."

Rely on emotion.

*(answers collected using a 7-item Likert scale from "Strongly Disagree" to "Strongly Agree")*

(page break)

"It is permissible to torture an innocent person if this would be necessary to provide information to prevent a bomb going off that would kill hundreds of people."

Rely on emotion.

*(answers collected using a 7-item Likert scale from "Strongly Disagree" to "Strongly Agree")*

(page break)

"It is just as wrong to fail to help someone as it is to actively harm them yourself."

Rely on emotion.

*(answers collected using a 7-item Likert scale from "Strongly Disagree" to "Strongly Agree")*

(page break)

"Sometimes it is morally necessary for innocent people to die as collateral damage—if more people are saved overall."

Rely on emotion.

*(answers collected using a 7-item Likert scale from "Strongly Disagree" to "Strongly Agree")*

(page break)

"It is morally wrong to keep money that one doesn't really need if one can donate it to causes that provide effective help to those who will benefit a great deal."

Rely on emotion.

*(answers collected using a 7-item Likert scale from "Strongly Disagree" to "Strongly Agree")*